\documentclass[12pt]{extarticle}
\usepackage{setspace}

\usepackage[T1]{fontenc}

\usepackage[latin1]{inputenc}
\usepackage{epsfig}
\usepackage[english,french]{babel}
\usepackage{color}
\usepackage{graphicx}

\usepackage{dcolumn}

\usepackage{moreverb}

\usepackage{amsmath,amssymb,amsfonts}

\begin{document}
\numberwithin{equation}{section}
\newcommand{\boxedeqn}[1]{%
  \[\fbox{%
      \addtolength{\linewidth}{-2\fboxsep}%
      \addtolength{\linewidth}{-2\fboxrule}%
      \begin{minipage}{\linewidth}%
      \begin{equation}#1\end{equation}%
      \end{minipage}%
    }\]%
}

%\boxedeqn{}

\newsavebox{\fmbox}
\newenvironment{fmpage}[1]
     {\begin{lrbox}{\fmbox}\begin{minipage}{#1}}
     {\end{minipage}\end{lrbox}\fbox{\usebox{\fmbox}}}

\raggedbottom
\onecolumn
%\pagestyle{headings}

%\begin{flushleft}
\begin{center}
\title*{{\LARGE{\textbf{Superintegrability and higher order polynomial algebras I}}}}
\end{center}
Ian Marquette
\newline
D\'epartement de physique et Centre de recherche math\'ematique,
Universit\'e de Montr\'eal,
\newline
C.P.6128, Succursale Centre-Ville, Montr\'eal, Qu\'ebec H3C 3J7,
Canada
\newline
ian.marquette@umontreal.ca
\newline
\newline
We present a method to obtain higher order integrals and polynomial algebras for two-dimensional superintegrable systems from creation and annihilation operators. All potentials with a second and a third order integrals of motion separable in Cartesian coordinates were studied. The integrals of motion of two of them do not generate a cubic algebra. We construct for these Hamiltonians a higher order polynomial algebra from the creation and annihilation operators. We obtain quintic and seventh order polynomial algebras. We give also for polynomial algebras of order 7 realizations in terms of deformed oscillator algebras. These realizations and finite dimensional unitary representations allow us to obtain the energy spectrum. 
\newline
\section{Introduction}
Over the years many articles were devoted to superintegrability [1-12]. In classical mechanics a Hamiltonian system with Hamiltonian H and integrals of motion $X_{a}$
\newline
\begin{equation}
H=\frac{1}{2}g_{ik}p_{i}p_{k}+V(\vec{x},\vec{p}),\quad X_{a}=f_{a}(\vec{x},\vec{p}),\quad a=1,..., n-1 \quad,
\end{equation}
\newline
is called completely integrable (or Liouville integrable) if it
allows n integrals of motion (including the Hamiltonian) that are
well defined functions on phase space, are in involution
$\{H,X_{a}\}_{p}=0$, $\{X_{a},X_{b}\}_{p}=0$, a,b=1,...,n-1 and
are functionally independent ($\{,\}_{p}$ is a Poisson bracket). A
system is superintegrable if it is integrable and allows further
integrals of motion $Y_{b}(\vec{x},\vec{p})$, $\{H,Y_{b}\}_{p}=0$,
b=1,...,k that are also well defined functions on phase
space and the integrals$\{H,X_{1},...,X_{n-1},Y_{1},...,Y_{k}\}$
are functionally independent. A system is maximally
superintegrable if the set contains 2n-1 functions such integrals. The integrals
$Y_{b}$ are not required to be in evolution with
$X_{1}$,...$X_{n-1}$, nor with each other. The same definitions apply in quantum mechanics but
$\{H,X_{a},Y_{b}\}$ are well defined quantum mechanical operators,
assumed to form an algebraically independent set. Superintegrable systems in classical and quantum mechanics
possess many properties. These systems appear to be important from the point of view of mathematics and physics. One
of their interesting property are their nonabelian algebraic structure generated by their integrals of motion. This algebraic structure can be a finite dimensional Lie algebra [1,2,3], a Kac-Moody algebra [13] or a polynomial algebra [14,15,16]. These polynomial algebras were related to deformed oscillator algebras and parafermionic algebras. Superintegrable systems are also related to systems studied in supersymmetric quantum mechanics [17-25]. Supersymmetry in quantum mechanics allows us to obtain the wave functions and the energy spectrum.
\newline
This article follows a series of articles [26,27,28,29,30,31] devoted to superintegrable systems in classical and quantum mechanics with third order integrals. In two-dimensional Euclidean space $E_{2}$ there are fourteen quantum systems with a second and a third order integrals [27]. These systems were investigated from the point of view of cubic algebras and supersymmetric quantum mechanics [30,31]. The Hamiltonians and the integrals of motion were of the following form
\newline
\begin{equation}
H=\frac{P_{x}^{2}}{2}+\frac{P_{y}^{2}}{2}+g_{1}(x)+g_{2}(y) \quad ,
\end{equation}
\begin{equation}
A=\frac{P_{x}^{2}}{2}-\frac{P_{y}^{2}}{2}+g_{1}(x)-g_{2}(y)\quad ,
\end{equation}
\begin{equation}
B=\sum_{i+j+k=3}A_{ijk}\{L_{3}^{i},p_{1}^{j}p_{2}^{k}\}+\{l_{1}(x,y),p_{1}\}+\{l_{2}(x,y),p_{2}\}\quad ,
\end{equation}
\newline
where $\{,\}$ is an anticommutator, $L_{3}=xP_{2}-yP_{1}$ is the angular
momentum and [H,A]=[H,B]=0. The constants $A_{ijk}$ and functions $l_{1}$ and $l_{2}$ are known [27].
\newline
We considered the most general cubic algebra generated by the integrals 
\newline
\begin{subequations}
\begin{equation}
[A,B]=C 
\end{equation}
\begin{equation}
[A,C]=\alpha A^{2} + \beta \{A,B\} + \gamma A + \delta B + \epsilon
\end{equation}
\begin{equation}
[B,C]=\mu A^{3} + \nu A^{2} - \beta B^{2} - \alpha \{A,B\} + \xi A -
\gamma B + \zeta \quad . 
\end{equation}
\end{subequations}
where $\alpha$, $\beta$, $\gamma$, $\delta$, $\epsilon$, $\xi$, $\nu$ and $\zeta$ are polynomials in H. We constructed realizations in terms of deformed oscillator algebras and found Fock type representations [29,30]. For twelve cases, the cubic algebra belongs in the following particular case
\newline
\begin{equation}
[A,B]=C,\quad  [A,C]=\delta B,\quad [B,C]=\mu A^{3} + \nu A^{2} + \xi A + \zeta \quad . 
\end{equation}
There are 6 irreducible (i.e. the third order integral is not a consequence of lower order one) quantum superintegrable Hamiltonians with a second and a third order integrals separable in Cartesian coordinates written with a rational function:
\newline
\newline
Potential 1. $V=\hbar^{2}[
\frac{x^{2}+y^{2}}{8a^{4}} +
\frac{1}{(x-a)^{2}}+\frac{1}{(x+a)^{2}}]$
\newline
Potential 2. $V=\frac{\omega^{2}}{2}(9x^{2} + y^{2})  $
\newline
Potential 3. $V=\frac{\omega^{2}}{2}(9x^{2} + y^{2})+\frac{\hbar^{2}}{y^{2}}    $
\newline
Potential 4. $V=\hbar^{2}[
\frac{9x^{2}+y^{2}}{8a^{4}} +
\frac{1}{(y-a)^{2}}+\frac{1}{(y+a)^{2}}]$
\newline
Potential 5. $V=\hbar^{2}(\frac{1}{8a^{4}}[(x^{2}+y^{2})+\frac{1}{y^{2}}+\frac{1}{(x+a)^{2}}+\frac{1}{(x-a)^{2}}
]  $
\newline
Potential 6. $V=\hbar^{2}[\frac{1}{8a^{4}}(x^{2}+y^{2})+\frac{1}{(y+a)^{2}}+\frac{1}{(y-a)^{2}}
+\frac{1}{(x+a)^{2}}+\frac{1}{(x-a)^{2}} ]   $ .
\newline
\newline
The integrals A,B,C respectively of order 2,3 and 4 of the Potential 5 and 6 do not generate a cubic algebra [30]. We studied these six potentials from the point of view of supersymmetric quantum mechanics (SUSYQM) and obtained the ladder operators, wave functions and energy spectrum.
\newline
Let us present the organization of this paper. In Section 2, we will show how we can generate higher order integrals and a polynomial algebras for two-dimensional Hamiltonians constructed from two one-dimensional Hamiltonians and their creation and annihilation operators. We present the polynomial algebra for two cases. In Section 3, we apply the results of Section 2 to the Smorodinsky-Winternitz potential and the Potential 5 and 6. The method presented in Section 2 can also be used to generate new superintegrable systems from known one-dimensional Hamiltonians for which creation and annihilation operators polynomial in momenta exist. We generalize the potential with a fourth Painlev\'e transcendent given by Eq.(1.10) and also construct its polynomial algebra with the results of Section 2. These results extend the number of known superintegrable systems involving the Painlev\'e transcendents. In Section 4, we give for a class of polynomial algebras of order 7 the realizations in terms of deformed oscillator algebras. In Section 5, we use the results of Section 3 and 4 to obtain the Fock type unitary representations and the corresponding energy spectrum of the Potential 5 and 6. 
\newline
\section{Polynomial algebras}
Let us consider a two-dimensional Hamiltonian separable in Cartesian coordinates
\newline
\begin{equation}
H(x,y,P_{x},P_{y})=H_{1}(x,P_{x})+H_{2}(y,P_{y}),
\end{equation}
\newline
for which creation and annihilation operators $A_{x}$, $A_{x}^{\dagger}$, $A_{y}$ and $A_{y}^{\dagger}$ (polynomials in momenta) exist. These operators satisfy the relations
\newline
\begin{equation}
[H_{1},A_{x}^{\dagger}]=\lambda_{x}A_{x}^{\dagger},\quad [H_{2},A_{y}^{\dagger}]=\lambda_{y}A_{y}^{\dagger}\quad .
\end{equation}
The following operators
\newline
\begin{equation}
f_{1}=A_{x}^{\dagger m}A_{y}^{n},\quad f_{2}=A_{x}^{m}A_{y}^{\dagger n}\quad ,
\end{equation}
commute with the Hamiltonian H given by Eq.(2.1)
\newline
\begin{equation}
[H,f_{1}]=[H,f_{2}]=0,
\end{equation}
if 
\begin{equation}
m\lambda_{x}-n\lambda_{y}=0, \quad m,n \in \mathbb{Z}^{+} \quad .
\end{equation}
Creation and annihilation operators allow to construct polynomial integrals of motion. The following sums are also polynomial integrals that commute with the Hamiltonian H
\begin{equation}
I_{1}=A_{x}^{\dagger m}A_{y}^{n}- A_{x}^{m}A_{y}^{\dagger n}, \quad I_{2}=A_{x}^{\dagger m}A_{y}^{n}+ A_{x}^{m}A_{y}^{\dagger n}\quad .
\end{equation}
The order of these integrals of motion depends of the order of the  creation and annihilation operators. The separation of variable in Cartesian coordinates implies the existence of a second order integral $K=H_{x}-H_{y}$. The creation and annihilation operators can also provide a method to determine the polynomial algebra of the two-dimensional superintegrable systems given by Eq.(2.1). We require that operators $A_{x}$ and $A_{x}^{\dagger}$ satisfy these further relations
\begin{equation}
[A_{x},A_{x}^{\dagger}]=P(H_{x})=Q(H_{x}+\lambda_{x})-Q(H_{x}),
\end{equation}
\begin{equation}
[A_{y},A_{y}^{\dagger}]=R(H_{y})=S(H_{y}+\lambda_{y})-S(H_{y}),
\end{equation}
where $P(H_{x})$ is of order $n_{x}$ and $Q(H_{x})$ is of order $n_{x}+1$ and $R(H_{y})$ is of order $n_{y}$ and $S(H_{y})$ is of order $n_{y}+1$. The algebras given by Eq.(2.7) and (2.8) are deformed oscillator algebras. They can be interpreted polynomial superalgebras i.e. $\{A_{x},A_{x}^{\dagger}\}=Q(H_{x} +\lambda_{x})+Q(H_{x})$.
\subsection{Case $\lambda_{x}=\lambda_{y}=\lambda$}
We consider the case
\begin{equation}
\lambda_{x}=\lambda_{y}=\lambda,
\end{equation}
and we take the following linear combination
\newline
\begin{equation}
A=2(H_{x}-H_{y}),\quad I_{1}=(A_{x}^{\dagger}A_{y}-A_{x}A_{y}^{\dagger}),
I_{2}=4\lambda (A_{x}^{\dagger}A_{y}+A_{x}A_{y}^{\dagger}) .
\end{equation}
\newline
We use the integrals given by Eq.(2.10) to construct polynomial algebras of the Hamiltonian given by Eq(2.1). 
\newline
\begin{equation}
[A,I_{1}]=I_{2},\quad [A,I_{2}]=16\lambda^{2}I_{1},
\end{equation}
\[[I_{1},I_{2}]=8\lambda(Q(\frac{1}{2}(H+\frac{1}{2}A))S(\frac{1}{2}(H-\frac{1}{2}A)+\lambda)\]
\[-Q(\frac{1}{2}(H+\frac{1}{2}A)+\lambda)S(\frac{1}{2}(H-\frac{1}{2}A)) .\]
The order of this polynomial algebra depends of the order of the polynomial Q and S. An example of such construction was used write the angular momentum algebra as two independent harmonic oscillators [32]. The construction of integrals of motion from creation and annihilation operators was discussed in earlier articles for specific examples: The two-dimensional harmonic oscillator [33], the three dimensional harmonic oscillator and the hydrogen atom [34], the Smorodinsky-Winternitz potentials [6,35] and the anisitropic oscillator in Ref.36.
\subsection{Case $\frac{\lambda_{x}}{\lambda_{y}}=\frac{1}{2}$}
We consider the case
\begin{equation}
2\lambda_{x}=\lambda_{y}=\lambda \quad ,
\end{equation}
and take the following integrals
\newline
\begin{equation}
A=2(H_{x}-H_{y}),\quad I_{1}=(A_{x}^{\dagger 2}A_{y}-A_{x}^{2}A_{y}^{\dagger}),
I_{2}=4 \lambda (A_{x}^{\dagger 2}A_{y}+A_{x}^{2}A_{y}^{\dagger}) .
\end{equation}
\newline
The polynomial algebra is thus
\newline
\begin{equation}
[A,I_{1}]=I_{2},\quad [A,I_{2}]=16 \lambda^{2} I_{1},
\end{equation}
\[[I_{1},I_{2}]=8 \lambda (Q(\frac{1}{2}(H+\frac{1}{2}A)-\lambda_{x})Q(\frac{1}{2}(H+\frac{1}{2}A))S(\frac{1}{2}(H-\frac{1}{2}A)+\lambda_{y})  \]
\[ -  Q(\frac{1}{2}(H+\frac{1}{2}A)+ 2\lambda_{x})Q(\frac{1}{2}(H+\frac{1}{2}A)+\lambda_{x})S(\frac{1}{2}(H-\frac{1}{2}A)).\]
The order of this polynomial algebra depends of the order of polynomials Q and S. Polynomial algebra given by Eq.(2.11) amd Eq.(2.14) have a similar structure as the cubic algebra given by Eq.(1.9).
\section{Applications}
There are two Smorodinsky-Winternitz potentials [6] that allow separation of variables of the Schrödinger equation in Cartesian coordinates: $V(x,y)=\frac{\omega}{2}(x^{2}+y^{2})+\frac{b}{x^{2}}+\frac{c}{y^{2}}$ and $V(x,y)=\frac{\omega^{2}}{2}(4x^{2}+y^{2}) + bx+\frac{b}{c^{2}}$. They are well known quadratically superintegrable systems and we apply the construction to these two systems. They have in the y axis the following creation operators
\newline
\begin{equation}
A_{x}^{\dagger}=-\frac{1}{4}(\frac{\hbar}{\omega}\frac{d^{2}}{dx^{2}}-2x \frac{d}{dx}+\frac{\omega}{\hbar}x^{2}-\frac{2b}{\omega\hbar x^{2}}-1) .
\end{equation}
It was first obtained in Ref. 6 and reobtained in a systematic study of systems with second order ladder operators [37].
The three other creation and annihilation operators have the same form. The polynomial $Q(H_{x})$ is given by
\begin{equation}
Q(H_{x})=\frac{1}{4\hbar^{2}\omega^{2}}H_{x}^{2}-\frac{1}{2\hbar\omega}H_{x}+(\frac{3}{16}-\frac{b}{2\hbar^{2}}),
\end{equation}
The polynomial $S(H_{y})$ is also given by the Eq.(3.2) (by replacing $H_{x}$ by $H_{y}$ and $b$ by $c$). We can form with $Q(H_{x})$, $S(H_{y})$, Eq.(2.10) and (2.11) the integrals and the polynomial algebra. In the two cases the polynomial algebra is a cubic algebra where the generators are second, third and fourth order operators. Hovewer, we can form a simpler algebraic structure for these Hamiltonians, a quadratic algebra where the generators are two second order and one third order operator [16]. The integrals obtained from the construction of Section 2 are not necessarily the integral of the lowest possible order. 
\subsection{Systems with a third order integral} 
We considered well known quadratically superintegrable systems. We will apply the method to Potential 1, 5 6 and present their integrals and polynomial algebras. The first system that we consider is the following [27,30]
\newline
\begin{equation}
H= \frac{1}{2}P_{x}^{2}+\frac{1}{2}P_{y}^{2}+\hbar^{2}( \frac{x^{2}+y^{2}}{8a^{4}} +
\frac{1}{(x-a)^{2}}+\frac{1}{(x+a)^{2}}) \quad  .
\end{equation}
The creation operators are given by
\begin{equation}
A_{x}^{\dagger}=\frac{\hbar^{2}}{4a^{2}}(-\frac{d}{dx} - \frac{1}{2a^{2}}x
+(\frac{1}{x-a}+\frac{1}{x+a}))(x-2a^{2}\frac{d}{dx})(\frac{d}{dx} - \frac{1}{2a^{2}}x +(\frac{1}{x-a}+\frac{1}{x+a})),
\end{equation}
\begin{equation}
A_{y}^{\dagger}=\frac{\hbar}{2a^{2}}(y-2a^{2}\frac{d}{dy})\quad .
\end{equation}
\newline
We construct the known integral B of order 3 [27,30] from Eq.(2.10) ($B=I'_{1}=\frac{-2a^{2}i}{\hbar}I_{1}$). We have $\lambda=\frac{\hbar^{2}}{2a^{2}}$. The deformed oscillator algebras in the x and y axis are given by Eq.(2.7) and (2.8) with the following expressions
\newline
\begin{equation}
Q(H_{x})=2H_{x}^{3}-\frac{7}{2}\frac{\hbar^{2}}{a^{2}}H_{x}^{2}+\frac{7\hbar^{4}}{8 a^{4}}H_{x}+\frac{15\hbar^{6}}{32a^{6}},
\end{equation}
\[S(H_{y})=2H_{y}-\frac{\hbar^{2}}{2a^{2}}. \]
\newline
We get from the Eq.(2.11) of the previous section and Eq.(3.6) the following cubic algebra that coincide with the one found in Ref. 30. 
\newline
\begin{equation}
[A,I'_{1}]=I'_{2} , \quad [A,I'_{2}]=\frac{4h^{4}}{a^{4}}I'_{1},
\end{equation}
\[ [I'_{1},I'_{2}]= -2\hbar^{2}A^{3} - 6\hbar^{2}A^{2}H + 8\hbar^{2}H^{3}
+ 6\frac{\hbar^{4}}{a^{2}}A^{2} + 8\frac{\hbar^{4}}{a^{2}}HA \]
\[-8\frac{\hbar^{4}}{a^{2}}H^{2} + 2\frac{\hbar^{6}}{a^{4}}A - 2\frac{\hbar^{6}}{a^{4}}H -
6\frac{\hbar^{8}}{a^{6}}. \]
The energy spectrum was calculated from the Fock type unitary representations [30]. In Section 4, we will extend this algebraic method of calculating the energy spectrum of superintegrable systems with a polynomial algebra of order seven.
\subsection{Potential 6} 
The next system that we consider is an Hamiltonian for which no polynomial algebra were found from the second and third order integrals of motion. We will show how the procedure of Section 2 will allow us to find a quintic algebra. The Hamiltonian
\newline
\begin{equation}
H= \frac{1}{2}P_{x}^{2}+\frac{1}{2}P_{y}^{2}+\hbar^{2}( \frac{x^{2}+y^{2}}{8a^{4}} +
\frac{1}{(x-a)^{2}}+\frac{1}{(x+a)^{2}}+\frac{1}{(y-a)^{2}}+\frac{1}{(y+a)^{2}}),
\end{equation}
has the following second order integral $A=H_{x}-H_{y}$ and third order integral
\newline
\begin{equation}
B=2L^{3}-3\alpha^{2}(\{L,P_{x}^{2}\}+\{L,P_{y}^{2}\})+\frac{\hbar^{2}}{4}\{(124y+3(\frac{y}{a^{2}})(x^{2}+y^{2})+24y\frac{(x^{2}-5y^{2})}{(y^{2}-a^{2})}
\end{equation}
\[-\frac{144yx^{2}}{x^{2}-a^{2}}+24y\frac{(3x^{2}-y^{2})(x^{2}+a^{2})}{(x^{2}-a^{2})^{2}}+48y\frac{(y^{2}-x^{2})(y^{2}+a^{2}}{(y^{2}-a^{2})},P_{x}\}\]
\[-\frac{\hbar^{2}}{4}\{(124x+3(\frac{x}{a^{2}})(y^{2}+x^{2})+24x\frac{(y^{2}-5x^{2})}{(x^{2}-a^{2})}
-\frac{144xy^{2}}{y^{2}-a^{2}}\]
\[+24x\frac{(3y^{2}-x^{2})(y^{2}+a^{2})}{(y^{2}-a^{2})^{2}}+48x\frac{(x^{2}-y^{2})(x^{2}+a^{2}}{(x^{2}-a^{2})},P_{y}\}.\]
\newline
The creation operators are given by Eq.(3.4) in the x and y axis (by replacing x by y). The polynomial algebras in the x and y axis are given by Eq.(2.7) and (2.8) with $Q(H_{x})$ and $S(H_{y})$ given by Eq.(3.6) (by replacing $H_{x}$ by $H_{y}$ for $S(H_{y})$). We have $\lambda=\frac{\hbar^{2}}{2a^{2}}$. The integrals of motion are given by Eq.(2.10) ($I'_{1}=\frac{-2a^{2}i}{\hbar}I_{1}$). Thus, we obtain with the Eq.(2.11) 
\newline
\begin{equation}
[A,I'_{1}]=I'_{2},\quad [A,I'_{2}]=\frac{4\hbar^{4}}{a^{4}}I'_{1}
\end{equation}
\[ [I'_{1},I'_{2}]=-\frac{3}{16}\hbar^{2}A^{5}+\frac{3}{2}\hbar^{2}A^{3}H^{2}-\frac{2\hbar^{4}}{a^{2}}A^{3}H-3\hbar^{2}AH^{4}
+\frac{8\hbar^{4}}{a^{2}}AH^{3}\]
\[+\frac{19\hbar^{6}}{8a^{4}}A^{3}-\frac{13\hbar^{6}}{2a^{4}}AH^{2}-\frac{99\hbar^{10}}{16a^{8}}A+\frac{6\hbar^{8}}{a^{6}}AH.\]
\newline
The integrals $I'_{1}$ is related to integrals A, B and C by 
\newline
\begin{equation}
I'_{1}=\frac{-1}{384\hbar^{2}}[A,C]+\frac{3\hbar^{2}}{32a^{4}}B.
\end{equation}
\subsection{Potential 5}
The Hamiltonian
\begin{equation}
H=\frac{1}{2}P_{x}^{2}+\frac{1}{2}P_{y}^{2}+\hbar^{2}(\frac{x^{2}+y^{2}}{8a^{4}}+\frac{1}{(x+a)^{2}}+\frac{1}{(x-a)^{2}}+\frac{1}{y^{2}})\quad ,
\end{equation}
has a quadratic $A=H_{x}-H_{y}$ and a cubic integrals
\newline
\begin{equation}
B=2L^{3}-3a^{2}\{L,P_{y}\}+\hbar^{2}\{\frac{3}{4a^{2}}-\frac{6y^{3}(x^{2}+a^{2})}{(x^{2}-a^{2})^{2}}-\frac{3(x^{2}-a^{2})}{y}-2y,P_{x}\}
\end{equation}
\[3\hbar^{2}\{x(\frac{(x^{2}-3a^{2})}{y^{2}}-\frac{(3y^{2}-8a^{2})}{12a^{2}}-\frac{2y^{2}}{x^{2}-a^{2}}+\frac{4y^{2}(x^{2}+a^{2})}{(x^{2}-a^{2})^{2}},P_{y}\}.\]
The integral A,B and their commutator C do not generate a cubic algebra. We will construct other integrals of motion from the creation and annihilation operators. The creation operators are given by Eq.(3.1)(by replacing x by y and $\omega$ by $a$) and (3.4). We have in the x and y axis polynomial algebras given by Eq.(2.7) and (2.8) with $\lambda=\frac{\hbar^{2}}{a^{2}}$ and
\begin{equation}
Q(H_{x})=2H_{x}^{3}-\frac{7}{2}\frac{\hbar^{2}}{a^{2}}H_{x}^{2}+\frac{7\hbar^{4}}{8 a^{4}}H_{x}+\frac{15\hbar^{6}}{32a^{6}},\quad 
\end{equation}
\[S(H_{y})=\frac{a^{4}}{\hbar^{4}}H_{y}^{2}-\frac{a^{2}}{\hbar^{2}}H_{y}-\frac{5}{16} \quad.\]
We obtain with the Eq.(2.14) and (3.31) the following polynomial algebra with integrals given by Eq.(2.10)( $I'_{1}=a^{2}I_{1}$)
\newline
\begin{equation}
[A,I'_{1}]=I'_{2},\quad [A,I'_{2}]=\frac{16\hbar^{4}}{a^{4}}I'_{1},
\end{equation}
\[[I'_{1},I'_{2}]=\frac{75\hbar^{14}}{64a^{10}}-\frac{275H\hbar^{12}}{64a^{8}}-\frac{3H^{2}\hbar^{10}}{16a^{6}}+\frac{261H^{3}\hbar^{8}}{16a^{4}}-\frac{75H^{4}\hbar^{6}}{4a^{2}}+\frac{15H^{5}\hbar^{4}}{4}\]
\[+3a^{2}H^{6}\hbar^{2}-\frac{a^{4}A^{7}}{64}-a^{4}H^{7}+A^{6}(\frac{7a^{2}\hbar^{2}}{64}-\frac{7a^{4}H}{64})+A^{5}(-\frac{3}{16}H^{2}a^{4}+\frac{9}{16}H\hbar^{2}a^{2}-\frac{25\hbar^{4}}{64})\]
\[+A^{4}(\frac{45\hbar^{6}}{64a^{2}}-\frac{85H\hbar^{4}}{64}+\frac{5}{16}a^{2}H^{2}\hbar^{2}+\frac{5a^{4}H^{3}}{16})+A^{3}(\frac{21\hbar^{8}}{64a^{4}}+\frac{5H\hbar^{6}}{8a^{2}}+\frac{15H^{2}\hbar^{4}}{8}-\frac{5}{2}a^{2}H^{3}\hbar^{2}\]
\[+\frac{5a^{4}H^{4}}{4})+A^{2}(-\frac{127\hbar^{10}}{64a^{6}}+\frac{239H\hbar^{8}}{64a^{4}}-\frac{85H^{2}\hbar^{6}}{8a^{2}}+\frac{95H^{3}\hbar^{4}}{8}-\frac{15}{4}a^{2}H^{4}\hbar^{2}+\frac{3a^{4}H^{5}}{4})\]
\[+A(\frac{5\hbar^{12}}{64a^{8}}-\frac{35H\hbar^{10}}{16a^{6}}+\frac{229H^{2}\hbar^{8}}{16a^{4}}-\frac{55H^{3}\hbar^{6}}{2a^{2}}+\frac{55H^{4}\hbar^{4}}{4}+a^{2}H^{5}\hbar^{2}-a^{4} H^{6})\quad . \]
The integral $I'_{1}$ is of order 7 and the integral $I'_{2}$ is of order 8.
\section{Realizations of polynomial algebras}
In the previous Section, we generated polynomial algebras of many systems. These algebras were cubic, quintic and seventh order algebras. In earlier articles it was demonstrated that quadratic [16] and cubic [30] algebras can be realized as deformed oscillator algebras [38]. Deformed oscillator algebras allow to construct Fock type representations and obtain the energy spectrum. We will show that we can construct similar realizations. We consider the following polynomial algebra of order seven
\begin{equation*}
[A,B]=C,\quad [A,C]=\delta B,
\end{equation*}
\begin{equation}
[B,C]=mA^{7}+nA^{6}+\mu A^{5}+\nu A^{4}+\alpha A^{3} +\beta A^{2} +\gamma A + \epsilon \quad .
\end{equation}
where $A$ and $B$ are integrals and thus commute with the Hamiltonian $H$. The structure constants $m$, $n$, $\mu$, $\nu$, $\alpha$, $\beta$, $\gamma$ and $\epsilon$ are polynomials of the Hamiltonian. We do not impose a order to these polynomials and only make the hypothesis that integrals generate an algebra of the form given by Eq.(4.1). The Casimir operator satisfies
\newline
\begin{equation}
[K,A]=[K,B]=[K,C]=0,
\end{equation}
and this implies
\newline
\begin{equation}
K=C^{2}-\delta B^{2}+\frac{m}{4}A^{8}+\frac{2}{7}nA^{7}+ (\frac{\mu}{3}+\frac{7}{6}\delta m)A^{6}+ (\frac{2}{5}\nu+ \delta n) A^{5}
\end{equation}
\[+(\frac{\alpha}{2}+\frac{5}{6}\delta \mu-\frac{7}{12}\delta^{2}m)A^{4}+(\frac{2}{3}\beta +\frac{2}{3}\delta
\nu-\frac{1}{3}\delta^{2}n)A^{3}\]
\[+(\frac{\delta\alpha}{2}-\frac{1}{6}\delta^{2}\mu +\gamma+\frac{1}{6}\delta^{3}m)A^{2}+(2\epsilon+\frac{1}{3}\delta \beta +\frac{1}{21}\delta^{3}n-\frac{\delta^{2}\nu}{15})A.\]
\newline
The order of the Casimir operator depends of the order of A and B. Ultimately, the Casimir operator is written in terms of the Hamiltonian. There is a realization in terms of deformed oscillator algebras of the form
\newline
\begin{equation}
A=\delta(N+u),\quad B=b^{\dagger}+b,
\end{equation}
where u is an arbitrary constant. Where $\{N,b,b^{\dagger}\}$ satisfy
\begin{equation}
[N,b]=-b,\quad [N,b^{\dagger}]=b^{\dagger},\quad bb^{\dagger}=\Phi(N+1),\quad b^{\dagger}b=\Phi(N).
\end{equation}
\newline
With the third relation of the seventh order algebra given by Eq.(4.1) and the Casimir operator given by Eq.(4.3) we find
\newline
\begin{equation}
\Phi(N)=\frac{m}{16}\delta^{3}(N+u)^{8}+(\frac{n\delta^{\frac{5}{2}}}{14}-\frac{m\delta^{3}}{4})(N+u)^{7}+               (\frac{\mu\delta^{2}}{12}+\frac{7}{24}m\delta^{3}-\frac{n\delta^{\frac{5}{2}}}{4})(N+u)^{6}
\end{equation}
\[+(\frac{\nu\delta^{\frac{3}{2}}}{10}-\frac{\mu\delta^{2}}{4}+\frac{1}{4}n\delta^{\frac{5}{2}})(N+u)^{5}
+(\frac{\alpha\delta}{8}+\frac{5\mu\delta^{2}}{24}-\frac{\nu\delta^{\frac{3}{2}}}{4}-\frac{7}{48}\delta^{3}m)(N+u)^{4}\]
\[+(\frac{\beta\delta^{\frac{1}{2}}}{6}+\frac{\nu\delta^{\frac{3}{2}}}{6}-\frac{\alpha\delta}{4}-\frac{1}{12}\delta^{\frac{5}{2}}n)(N+u)^{3}+(\frac{\delta\alpha}{8}-\frac{\delta^{2}\mu}{24}+\frac{\gamma}{4}-\frac{\beta\delta^{\frac{1}{2}}}{4}+\frac{1}{24}\delta^{3}m)(N+u)^{2}\]
\[+(\frac{\epsilon}{2\delta^{\frac{1}{2}}}+\frac{\delta^{\frac{1}{2}}\beta}{12}-\frac{\gamma}{4}-\frac{1}{84}\delta^{\frac{5}{2}}n -\frac{1}{60}\delta^{\frac{3}{2}}\nu)(N+u)-\frac{\epsilon}{4\delta^{\frac{1}{2}}}-\frac{K}{4\delta}.\]
To obtain unitary representations we should impose three constraints on the structure function
\newline
\begin{equation}
\Phi(p+1,u_{i},k)=0, \quad \Phi(0,u,k)=0,\quad \phi(x)>0, \quad \forall \quad x>0 \quad .
\end{equation}
\section{Potential 5 and 6}
The Eq.(4.6) gives the structure function in terms of the parafermionic number $N$ and the structure constants. The three conditions given by Eq.(4.7) provide a method to obtain the energy spectrum. From the results of Section 4 and 3, we can found the unitary representations and the corresponding energy spectrum for the Potential 5 and 6.
\subsection{Potential 6 and quintic algebras}
The algebra of Potential 6 is given by the Eq.(3.10) is a particular case of the one given by the Eq.(4.1). The structure constant are
\newline
\begin{equation}
\delta =\frac{4 \hbar^{4}}{a^{4}}, \quad \mu =-\frac{3\hbar^{2}}{16},\quad \nu=\beta=\epsilon=0,
\end{equation}
\[\alpha=\frac{3}{2}\hbar^{2}H^{2}+\frac{2\hbar^{4}}{a^{2}}H+\frac{19\hbar^{6}}{8a^{4}},\quad   \gamma=-3\hbar^{2}H^{4}+\frac{8\hbar^{4}}{a^{2}}H^{3}-\frac{13\hbar^{6}}{2a^{4}}H^{2}+\frac{6\hbar^{8}}{a^{6}}H-\frac{99\hbar^{10}}{16a^{8}}.\]
\newline
This quintic algebra is generated by integrals A, $I'_{1}$ and $I'_{2}$ respectively of order 2, 5 and 6.
We can write the Casimir operator given by Eq.(4.3) as a polynomial of the Hamiltonian only
\newline
\begin{equation}
K=-4\hbar^{2}H^{6}+\frac{16\hbar^{4}}{a^{2}}H^{5}-\frac{5\hbar^{6}}{a^{4}}H^{4}-\frac{40\hbar^{8}}{a^{6}}H^{3}+\frac{141\hbar^{10}}{4a^{8}}H^{2}+\frac{9\hbar^{12}}{a^{10}}H-\frac{135\hbar^{14}}{16a^{12}}.
\end{equation}
We can found with the Eq.(4.6) the structure function and factorize it in the following way
\newline
\begin{equation}
\Phi(x)=\frac{-\hbar^{10}}{4a^{8}}((x+u)-(\frac{a^{2}E}{\hbar^{2}}-\frac{3}{2}))((x+u)-(\frac{-a^{2}E}{\hbar^{2}}-\frac{1}{2}))
\end{equation}
\[ ((x+u)-(\frac{a^{2}E}{\hbar^{2}}-\frac{1}{2}))((x+u)-(\frac{-a^{2}E}{\hbar^{2}}+\frac{3}{2}))\]
\[ ((x+u)-(\frac{a^{2}E}{\hbar^{2}}+\frac{3}{2}))((x+u)-(\frac{-a^{2}E}{\hbar^{2}}+\frac{5}{2})).\]
To obtain unitary representations we should impose three constraints given by Eq.(4.7). There are four solutions for $a=ia_{0}. \in \mathbb{R}$. Let us present these unitary representations with the corresponding constant u and energy spectrum:
\newline
Case with $u_{1}=\frac{a^{2}E}{\hbar^{2}}+\frac{3}{2}$
\begin{equation}
E_{1}=\frac{\hbar^{2}(p+3)}{2a_{0}^{2}}, 
\end{equation}
\begin{equation}
\Phi_{1}(x)=\frac{\hbar^{10}}{4a_{0}^{8}}x(x+2)(x+3)(p+4-x)(p+3-x)(p+1-x).
\end{equation}
Case with $u_{2}=\frac{a^{2}E}{\hbar^{2}}-\frac{1}{2}$
\begin{equation}
E_{2}=\frac{\hbar^{2}(p+1)}{2a_{0}^{2}},
\end{equation}
\begin{equation}
\Phi_{3}(x)=\frac{\hbar^{10}}{4a_{0}^{8}}x(x-2)(p+4-x)(p+3-x)(p+1-x),p=0,1.
\end{equation}
Case with $u_{3}=\frac{a^{2}E}{\hbar^{2}}-\frac{3}{2}$
\begin{equation}
E_{3}=\frac{\hbar^{2}(p)}{2a_{0}^{2}}, 
\end{equation}
\begin{equation}
\Phi_{3}(x)=\frac{\hbar^{10}}{4a_{0}^{8}}x(x-3)(p+1-x)(p+3-x)(p+4-x),p=0,1,2.
\end{equation}
\begin{equation}
E_{4}=\frac{\hbar^{2}(p-3)}{2a_{0}^{2}}, 
\end{equation}
\begin{equation}
\Phi_{4}(x)=\frac{\hbar^{10}}{4a_{0}^{8}}x(x-3)(p+1-x)(p-2-x)(p-x),p=0,
\end{equation}
We must also exclude spurious states with other conditions. One of them consists in $E \geq min \quad V$. We have unitary representations valid only for p=0, p=0,1 and p=0,1,2. Such solutions were also found in the context of cubic algebras.
This phenomenon is related to zero modes, singlet state, doublet states and higher order supersymmetric quantum mechanics [30,31]. A singlet state is annihilated by the annhilation and creation operators. The energy spectrum is confirmed by the results obtained from supersymmetric quantum mechanics.
\newline
There is one solution for the case $a \in \mathbb{R}$ with $u=\frac{-a^{2}E}{\hbar^{2}}+\frac{5}{2}$. The unitary representation is
\begin{equation}
E_{1}=\frac{\hbar^{2}(p+5)}{2a^{2}}, \quad p \geq 3,
\end{equation}
\begin{equation}
\Phi_{1}(x)=\frac{\hbar^{10}}{4a^{8}}x(p+1-x)(x+3)(p+4-x)(p+2-x).
\end{equation}
\subsection{Potential 5 and polynomial algebras of order 7}
The algebra of the Potential 5 is given by (3.15). This case belongs to the one given by Eq.(4.1). The structure constants are obtained by comparing the Eq.(3.15) and (4.1).
\newline
This seventh order algebra is generated by integrals A, $I'_{1}$ and $I'_{2}$ respectively  of order 2, 7 and 8. The Casimir operator given by Eq.(4.3) can be written as a function of the Hamiltonian only
\newline
\begin{equation}
K=a^{4}H^{8}-4a^{2}\hbar^{2}H^{7}+3\hbar^{4}H^{6}+\frac{15\hbar^{6}}{a^{2}}H^{5}-\frac{453\hbar^{8}}{8a^{4}}H^{4}
\end{equation}
\[+\frac{261\hbar^{10}}{4a^{6}}H^{3}-\frac{133\hbar^{12}}{16a^{8}}H^{2}-\frac{275\hbar^{14}}{16a^{10}}H+\frac{1425\hbar^{16}}{256a^{12}}.\]
The Eq.(4.6) give us the structure function and we can factorize it in the following way
\newline
\begin{equation}
\Phi(x)=(\frac{4\hbar^{12}}{a^{8}})(x+u-(-\frac{1}{4}-\frac{a^{2}E}{2\hbar^{2}}))(x+u-(-\frac{1}{4}+\frac{a^{2}E}{2\hbar^{2}}))
\end{equation}
\[  (x+u-(\frac{1}{4}-\frac{a^{2}E}{2\hbar^{2}}))(x+u-(\frac{3}{4}-\frac{a^{2}E}{2\hbar^{2}}))  (x+u-(\frac{5}{4}-\frac{a^{2}E}{2\hbar^{2}}))(x+u-(\frac{5}{4}-\frac{a^{2}E}{2\hbar^{2}})) \]
\[  (x+u-(\frac{5}{4}+\frac{a^{2}E}{2\hbar^{2}}))(x+u-(\frac{7}{4}-\frac{a^{2}E}{2\hbar^{2}})) . \]
\newline
Let us present the solutions for the case $a=a_{0}i, a_{0} \in \mathbb{R}$.
\newline
There is two solutions for $u=\frac{5}{4}+\frac{a^{2}E}{2\hbar^{2}}$ 
\newline
\begin{equation}
E_{1}=\frac{\hbar^{2}(5+2p)}{2a_{0}^{2}},\quad \Phi_{1}(x)=(\frac{\hbar^{12}}{4a_{0}^{8}})x(p+1-x)(2p+3-2x)(2p+5-2x)^{2}
\end{equation}
\[(p+2-x)(p+3-x)(3+2x).\]
\begin{equation}
E_{2}=\frac{\hbar^{2}(1+p)}{a_{0}^{2}},\quad \Phi_{2}(x)=(\frac{\hbar^{12}}{4a_{0}^{8}})(3+2p-2x)(5+2p-2x)^{2}
\end{equation}
\[(p+1-x)(p+2-x)(p+3-x)(-3+2x)\]
The representation $\Phi_{2}(x)$ is valid for p=0. We confirmed these energy levels with the results obtained using the separability in Cartesian coordinates and the SUSYQM.
\newline
There is one solution for the case $a \in \mathbb{R}$. For $u=2-\frac{a^{2}E}{\hbar^{2}}$ we get 
\newline
\begin{equation}
\Phi_{1}(x)=(\frac{\hbar^{12}}{4a^{8}})x(p+1-x)(x+\frac{1}{2})^{2}(x+\frac{3}{2})(x+2)(p+\frac{5}{2}-x),
\end{equation}
\begin{equation}
E_{1}=\frac{\hbar^{2}(p+3)}{a^{2}}.
\end{equation} 
\section{Conclusion}
In this article, we have showed how we can construct integrals of motion for two-dimensional Hamiltonians of the form given by Eq.(2.1) from the creation and annihilation operators of one-dimensional Hamiltonians $H_{x}$ and $H_{y}$. We construct from these integrals of motion higher order polynomial algebras. We presented the algebras for two cases: $\lambda_{x}=\lambda_{y}$ and $2\lambda_{x}=\lambda_{y}$. These polynomial algebra are given by the Eq.(2.11) and(2.14). These results, we explained the particular form of the cubic algebra obtained in Ref. 30 and given by Eq.(1.6).
\newline
One important result of this article is the construction of higher polynomial algebra for the Potential 5 and 6. They are respectively seventh order and quintic algebras. The quantum case appears to be more richer than the classical one in terms of algebraic structures. The integrals of motion of all classical superintegrable systems with a second and a third order integral in $E_{2}$ generate cubic Poisson algebra [28].
\newline
We have also studied the realization in terms of deformed oscillator algebras of a class of polynomial algebras of the seventh order. These results allowed us to obtain the structure function for the Potential 5 and 6 and to obtain unitary representations with their corresponding energy spectrum. We corobored these results with those obtained by the means of supersymmetric quantum mechanics [30].
\newline
These results can be generalized in higher dimensions. The method of the Section 2 can also be used to generate new superintegrable systems from known one-dimensional systems. 
\newline
\newline
\textbf{Acknowledgments} The research of I.M. was supported by a postdoctoral
research fellowship from FQRNT of Quebec. The author thanks P.Winternitz for very helpful comments and discussions.

\section{\textbf{References}}

1. V.Fock, Z.Phys. 98, 145-154 (1935).
\newline
2. V.Bargmann, Z.Phys. 99, 576-582 (1936).
\newline
3. J.M.Jauch and E.L.Hill, Phys.Rev. 57, 641-645 (1940).
\newline
4. M.Moshinsky and Yu.F.Smirnov, The Harmonic Oscillator In Modern
Physics, (Harwood, Amsterdam, 1966).
\newline
5. J.Fris, V.Mandrosov, Ya.A.Smorodinsky, M.Uhlir and P.Winternitz,  Phys.Lett. 16, 354-356 (1965).
\newline
6. P.Winternitz, Ya.A.Smorodinsky, M.Uhlir and I.Fris, Yad.Fiz. 4,
625-635 (1966). (English translation in Sov. J.Nucl.Phys. 4,
444-450 (1967)).
7. A.Makarov, Kh. Valiev, Ya.A.Smorodinsky and P.Winternitz, Nuovo Cim. A52, 1061-1084 (1967).
\newline
8. N.W.Evans, Phys.Rev. A41, 5666-5676 (1990), J.Math.Phys. 32,
3369-3375 (1991).
\newline
9. E.G.Kalnins, J.M.Kress, W.Miller Jr and P.Winternitz,
 J.Math.Phys. 44 (12) 5811-5848 (2003).
\newline
10. E.G.Kalnins, W.Miller Jr and G.S.Pogosyan, J.Math.Phys. A34,
4705-4720 (2001).
\newline
11. E.G.Kalnins, J.M.Kress and W.Miller Jr, J.Math.Phys. 46,
053509 (2005),46, 053510 (2005), 46, 103507 (2005), 47, 043514
(2006), 47, 043514 (2006).
\newline
12. E.G.Kalnins, W.Miller Jr and G.S.Pogosyan, J.Math.Phys. 47,
033502.1-30 (2006), 48, 023503.1-20 (2007).
\newline
13. J.Daboul, P.Slodowy et C.Daboul, Phys.Lett. B317, 321-328 (1993)
\newline
14. Ya.I.Granovsky, A.S.Zhedanov et I.M.Lutzenko, Ann. Phys. (New York) 217, 1, 1-20 (1992).
\newline
15. P.L\'etourneau et L.Vinet,  Ann. Phys. (New York) 243, 144-168 (1995).
\newline
16. C.Daskaloyannis, J.Math. Phys. 42, 1100-1119 (2001).
\newline
17. E.Witten, Nucl.Phys. B188, 513 (1981); E.Witten, Nucl.Phys. B202, 253-316 (1982).
\newline
18. L.Gendenshtein, JETP Lett., 38, 356 (1983).
\newline
19. B.Mielnik, J. Math. Phys. 25, 12 (1984).
\newline
20. G.Junker, Supersymmetric Methods in Quantum and Statistical Physics, Springer, New York, (1995).
\newline
21. M.S.Plyushchay, Annals Phys. (N.Y.) 245, 339 (1996); M.Plyushchay, Int.J.Mod.Phys. A15, 3679 (2000); F.Correa and M.Plyushchay,  Annals Phys. 322, 2493 (2007).
\newline
22. F.Cooper, A.Khare et U.Sukhatme, Phys.Rept. 251, 267-385 (1995).
\newline
23. A.Khare and R.K.Bhaduri, Am.J.Phys 62, 1008-1014 (1994)
\newline
24. I.F.Marquez, J.Negro and L.M.Nieto, J.Phys. A:Math. Gen. 31, 4115 (1998).
\newline
25. D.J.Fernandez C,V.Hussin and L.M.Nieto,  J.Phys.A:Math. Gen. 27, 3547 (1994);D.J.Fern\'andez, V.Hussin and O.Rosas-Ortiz,  J.Phys.A: Math.Theor. 40 6491-6511 (2007)
\newline
26. S.Gravel and P.Winternitz, J.Math.Phys. 43 (12), 5902 (2002).
\newline
27. S.Gravel, J.Math.Phys. 45 (3), 1003-1019 (2004).
\newline
28. I.Marquette and P.Winternitz, J.Math.Phys. 48 (1) 012902
(2007).
\newline
29. I.Marquette and P.Winternitz, J. Phys. A: Math. Theor. 41, 304031 (2008).
\newline
30. I.Marquette, J. Math. Phys. 50, 012101 (2009).
\newline
31. I.Marquette, J.Math.Phys. 50 095202 (2009).
\newline
32. J.Schwinger, AEC Report NYO-3071 (1952).
\newline
33. V.A.Dulock and H.V.McIntosh, Am. J.Phys. 33, 109 (1965).
\newline
34. A.Cisneros and H.V.McIntosh, J.Math.Phys. 11 3 870 (1970).
\newline
35. V.Sunilkumar, arXiv:math-ph/0203047 (2002).
\newline
36. D.Bonatsos, C.Daskaloyannis, Prog.Part.Nucl.Phys. 43, 537 (1999).
\newline
37. C.P.Boyer and W.Miller Jr., J.Math.Phys. 15, 9 (1974).
\newline
38. C.Daskaloyannis, J.Phys.A: Math.Gen 24, L789-L794 (1991).

%\end{flushleft}
\end{document}